# An acoustic metamaterial lens for acoustic point-to-point communication in air

Fei Sun[1], Shuwei Guo[1], Borui Li[1], Yichao Liu[1], and Sailing He[1, 2]

1 Centre for Optical and Electromagnetic Research, Zhejiang Provincial Key Laboratory for Sensing Technologies, JORCEP, East Building #5, Zhejiang University, Hangzhou 310058, China

2 Department of Electromagnetic Engineering, School of Electrical Engineering, Royal Institute of Technology (KTH), S-100 44 Stockholm, Sweden

**Abstract**

Acoustic metamaterials have become a novel and effective way to control sound waves and design acoustic devices. In this study, we design a 3D acoustic metamaterial lens (AML) to achieve point-to-point acoustic communication in air: any acoustic source (i.e. a speaker) in air enclosed by such an AML can produce an acoustic image where the acoustic wave is focused (i.e. the field intensity is at a maximum, and the listener can receive the information), while the acoustic field at other spatial positions is low enough that listeners can hear almost nothing. Unlike a conventional elliptical reflective mirror, the acoustic source can be moved around inside our proposed AML. Numerical simulations are given to verify the performance of the proposed AML.

**Introduction**

Acoustic metamaterials are made by artificial units on the sub-wavelength orders, which can manipulate the propagation of acoustic waves in an unprecedented way [1-3]. Many novel acoustic phenomena and devices have been proposed, and later experimentally demonstrated by acoustic metamaterials, including negative mass density and modulus [4-6], anisotropic mass density [7], acoustic beam shifters [8], acoustic cloaking [9-13], acoustic super-resolution imaging [14, 15], acoustic wave rotators [16], acoustic absorbers [17, 18], acoustic rainbow trapping [19, 20], acoustic translational illusion [21], etc.

Point-to-point communication is a relatively safe communication method. Point-to-point communication can be achieved easily through lightwaves or microwaves (e.g. laser communication systems [22], mobile phones, and telephone communication). However, it is challenging to achieve a point-to-point communication using acoustic waves due to the diffraction of acoustic waves (e.g. when one speaker is talking, many audiences around him can receive the acoustic information). Acoustic point-to-point communication often requires conversion of the acoustic signal to other kinds of signals (e.g. an electrical signal, in the case of cell phones), achieving point-to-point transmission with the help of the other signal, and finally converting it back to the acoustic signal. Can we create a special environment in which point-to-point acoustic communication can be achieved directly, without any conversion to and from other kinds of signal? In this study, we try to achieve such an effect by enclosing a circular region of air with an acoustic metamaterial lens (AML). The basic idea is shown in Fig. 1. The AML (colored green), filled with the designed inhomogeneous media, is set around the room (i.e. a circular air region (colored white)). The red small circle in the air represents the speaker, while the other small circles in the air all represent listeners. Of all the possible locations of the speaker in the room, there is only one fixed spatial location (colored brown) where a listener can receive the acoustic signal well (i.e. the intensity of the acoustic field here is large enough to be recognized). Listeners in other spatial locations (colored blue) cannot receive the information, as the intensity of the acoustic wave is below the human ear hearing range (e.g. below 20*dB*) in these regions. The key to achieve such an effect is to design the AML around the room.

**Method**

We will first theoretically show what types of inhomogeneous acoustic medium can achieve such an effect by making an analogy to the design of inhomogeneous media for electromagnetic waves and then design an acoustic metamaterial structure (i.e. layered media composed of air, petroleum, and hydrogen with a gradual filling factor) based on effective medium theory to realize such an acoustic lens. The propagation of a time-harmonic acoustic pressure in a source-free space obeys the Helmholtz equation [23]:

$$\bar{\nabla}^2 p(\bar{r}) + \frac{\rho(\bar{r})\omega^2}{\kappa(\bar{r})} p(\bar{r}) = 0, \qquad (1)$$

where $p$ is the acoustic pressure, $\rho$ is the mass density, $\omega$ is the angular frequency, and $\kappa$ is the modulus. The speed of the acoustic wave can be defined by:

$$v = \sqrt{\frac{\kappa(\bar{r})}{\rho(\bar{r})}}. \qquad (2)$$

We can rewrite Eq. (1) by introducing the speed of the acoustic wave in Eq. (2):

$$\overline{\nabla}^2 p(\overline{r}) + \frac{\omega^2}{v^2} p(\overline{r}) = 0. \tag{3}$$

Considering the propagation of electromagnetic waves in inhomogeneous media:

$$\overline{\nabla}^2 \psi(\overline{r}) + n^2 \frac{\omega^2}{c^2} \psi(\overline{r}) = 0, \tag{4}$$

where $\psi$ is the electric field, $n$ is the refractive index of the medium, and $c$ is speed of light in vacuum. By analogizing the electromagnetic wave equation (Eq. (4)) and the acoustic wave equation (Eq. (3)), we can also introduce the effective refractive index for an acoustic wave:

$$n_{eff} = \frac{v_0}{v}. \tag{5}$$

where $v_0$ and $v$ are the speeds of the acoustic wave in the background medium (e.g. air) and in another medium filled into the background medium (e.g. acoustic metamaterials filled in air). Since the forms of the acoustic wave equation and the scalar electromagnetic wave equation are the same, we would expect that for some inhomogeneous acoustic media whose effective refractive indices vary, much like a special lens that works for electromagnetic waves, their function on acoustic waves should be the same as their function on electromagnetic waves. The inside-out Eaton lens can achieve point-to-point imaging for electromagnetic waves [24-26]. We can design an acoustic medium whose effective refractive index fulfills the inside-out Eaton lens' profile [26]:

$$n_{eff} = \begin{cases} 1, r \leq a \\ \sqrt{\dfrac{2a}{r} - 1}, a < r \leq 2a \end{cases}. \tag{6}$$

Combining Eqs. (2), (5) and (6), the modulus and mass density of the acoustic inside-out Eaton lens should be:

$$\sqrt{\frac{\kappa(r)}{\rho(r)}} = \begin{cases} v_0, r \leq a \\ \dfrac{v_0}{\sqrt{\dfrac{2a}{r}-1}}, a < r \leq 2a \end{cases}. \qquad (7)$$

Now we have obtained the required medium (i.e. Eq. (7)) with an inhomogeneous ratio of the modulus and mass density to achieve point-to-point acoustic communication in Fig. 1. $v_0$ is the acoustic speed in air and $r$ is the radial distance from the center of the inside-out Eaton lens. $a$ is the radius of the inner air region and $2a$ is the outer radius of the green annular region in Fig. 1. Note that there is a rigid boundary at $r=2a$. Another requirement on the modulus and mass density of this aoucstic lens is the impedance match condition (i.e. the impedance of the lens should match the air's impedance):

$$\rho(r)v(r) = \rho_0 v_0 \qquad (8)$$

Combing Eqs. (5), (6) and (8), we can obtain the required mass density distribution of the impedance-matched, acoustic inside-out Eaton lens:

$$\rho(r) = \begin{cases} \rho_0, r \leq a \\ \rho_0 \sqrt{\dfrac{2a}{r}-1}, a < r \leq 2a \end{cases}. \qquad (9)$$

Putting Eq. (9) into Eq. (7), we can obtain the required modulus of the impedance-matched acoustic inside-out Eaton lens:

$$\kappa(r) = \begin{cases} \rho_0 v_0^2, r \leq a \\ \dfrac{\rho_0 v_0^2}{\sqrt{\dfrac{2a}{r}-1}}, a < r \leq 2a \end{cases}. \qquad (10)$$

We use the commercial software COMSOL Multiphysics 5.0, based on the finite element method, to verify the performance of the acoustic inside-out Eaton lens. The performance of the impedance-matched acoustic inside-out Eaton lens, whose mass density and modulus are given in Eqs. (9) and (10), is shown in Fig. 2. If the green annular medium ($a<r<2a$) is filled by acoustic materials with the modulus and mass density given in Eqs. (9) and (10), and the white circle region ($r<a$) is filled with air, any vibration source at ($x_0$, $y_0$) in air can produce a foucsed acoustic image at $(-x_0, -y_0)$. In other words, we can achieve point-to-point acoustic communication between $(x_0, y_0)$ and $(-x_0, -y_0)$ by the medium given in Eqs. (9) and (10).

**Realiation Design**

Next, we will show how to use layered media (air, petroleum, and hydrogen) to realize the required medium in Eqs. (9) and (10). Fig. 3(a) is a schematic diagram of our AML composed by air (colored blue), petroleum (colored red), and hydrogen (colored yellow). Based on the effective medium theory [27, 28], the effective mass density and modulus of the layered media (see Fig. 3(b)) can be given by:

$$\begin{cases} \dfrac{1}{\rho_\parallel} = \dfrac{f_A}{\rho_A} + \dfrac{f_B}{\rho_B} + \dfrac{f_C}{\rho_C} \\ \rho_\perp = f_A \rho_A + f_B \rho_B + f_C \rho_C, \\ \dfrac{1}{\kappa} = \dfrac{f_A}{\kappa_A} + \dfrac{f_B}{\kappa_B} + \dfrac{f_C}{\kappa_C} \end{cases} \quad (11)$$

where $f_A = d_A/d$, $f_B = d_B/d$ and $f_C = d_C/d$ are filling factors ($f_A + f_B + f_C = 1$, $0 < f_i < 1$, i=A, B, C). We choose air, petroleum, and hydrogen as media A, B and C, respectively. The modulus of these media are $\kappa_A = v_0^2 \rho_0 \sim 1.42 \times 10^5$ Pa, $\kappa_B = 1.2382 \times 10^9$ Pa and $\kappa_C = 0.95 \times 10^9$ Pa. The mass density of these mediua are $\rho_A = 1.21$ kg/m$^3$, $\rho_B = 700$ kg/m$^3$, and $\rho_C = 0.09$ kg/m$^3$. Note that we only consider 2D case (acoustic waves are within the *x-y* plane). Therefore, only mass denstiy wihtin the *x-y* plane works. In this case, Eq. (11) can be rewritten as:

$$\begin{cases} \dfrac{1}{\rho_\parallel} = \dfrac{f_A}{\rho_A} + \dfrac{f_B}{\rho_B} + \dfrac{f_C}{\rho_C} \\ \dfrac{1}{\kappa} = \dfrac{f_A}{\kappa_A} + \dfrac{f_B}{\kappa_B} + \dfrac{f_C}{\kappa_C} \\ f_A + f_B + f_C = 1 \end{cases} . \quad (12)$$

In order to realize the AML, we need first to make some simplifications: the first simplification is to the outmost boundary is chosen at $r=1.85a$ (we set a rigid boundary at $r=1.85a$) not $r=2a$ due to infinite the infinite modulus at $r=2a$ in Eq. (10), which cannot be realized by any materials). Numerical simulations show that this simplification does not have obvious influence on the performance of the AML (the field distribution is still almost like Fig.2; we do not show them here). The second simplification is to discretize the whole lens to seventeen different layers (see Fig. 3(a)), in each of which the effective mass density and modulus are constant. From the inner layer to the outer layer, we arrange a layer number, i.e. No.1, 2, …, 17, and the radius of each layer's inner and outer boundaries are shown in Tab. 1.The third simplification is to set filling factors $f_A=0$, $f_B=0.8$, and $f_C=0.2$ at the last layer (No. 17). Note that the accurate calculated filling factors from Eqs. (9, 10, 12) should be $f_B=0.8212$, and $f_C=0.2287$ in which the condittion $f_A+f_B+f_C=1$ and $0<f_A<1$ cannot be satisified at the same time. The simplificated AML is composed of seventeen layers of annular layered media with different filling factors along the $z$ direction (see Fig. 3(a)). The detailed parameters of each layer are given in Tab. 1. The performance of the simplificated AML is shown in Fig. 4. As shown in Fig. 4, the simplificated AML can still provide good point-to-point imaging.

**Discussion and Conclusion**

Our AML is different from the echo wall, which can achieve long-distance acoustic communication with extremely small attenuation around the wall. An echo wall cannot achieve point-to-point acoustic communication (i.e. other listeners around the wall can also receive the acoustic signal). For an elliptical reflective mirror, only two focal points can achieve point-to-point acoustic communication. However, there are infinitely many spatial locations where such an effect can be achieved in our AML. Even if the speaker is moving, such an acoustic point-to-point relation still holds. Note that our AML cannot achieve a super-resolution acoustic imaging, as positive inhomogeneous media cannot amplify/convert the evanescence waves like other lenses [14, 15].

By analogizing the wave equations between acoustic waves and electromagnetic waves, we design an inhomogeneous acoustic lens, i.e. an inside-out Eaton lens. Numerical simulations show its point-to-point imaging ability, which can be utilized for point-to-point acoustic communication. After some simplifications, we design an AML composed of layered air, petroleum, and hydrogen to realize the acoustic inside-out Eaton lens based on the effective medium theory, the performance of which is also verified by numerical simulations. The proposed lens will have many potential applications in acoustic point-to-point communications, acoustic wave focusing, etc. The proposed theoretical method (i.e. an acousto-optic analogy) can also be utilized to design many other novel acoustic devices.

**Acknowledgment**

This work is partially supported by the National Natural Science Foundation of China (Nos. 11604292, 91233208 and 60990322), the National High Technology Research and Development Program (863 Program) of China (No. 2012AA030402), the fundamental research funds for the central universities (No. 2017FZA5001), the Program of Zhejiang Leading Team of Science and Technology Innovation, the Postdoctoral Science Foundation of China (No. 2017T100430), the Preferred Postdoctoral Research Project Funded by Zhejiang Province (No. BSH1301016) and Swedish VR grant (# 621-2011-4620).

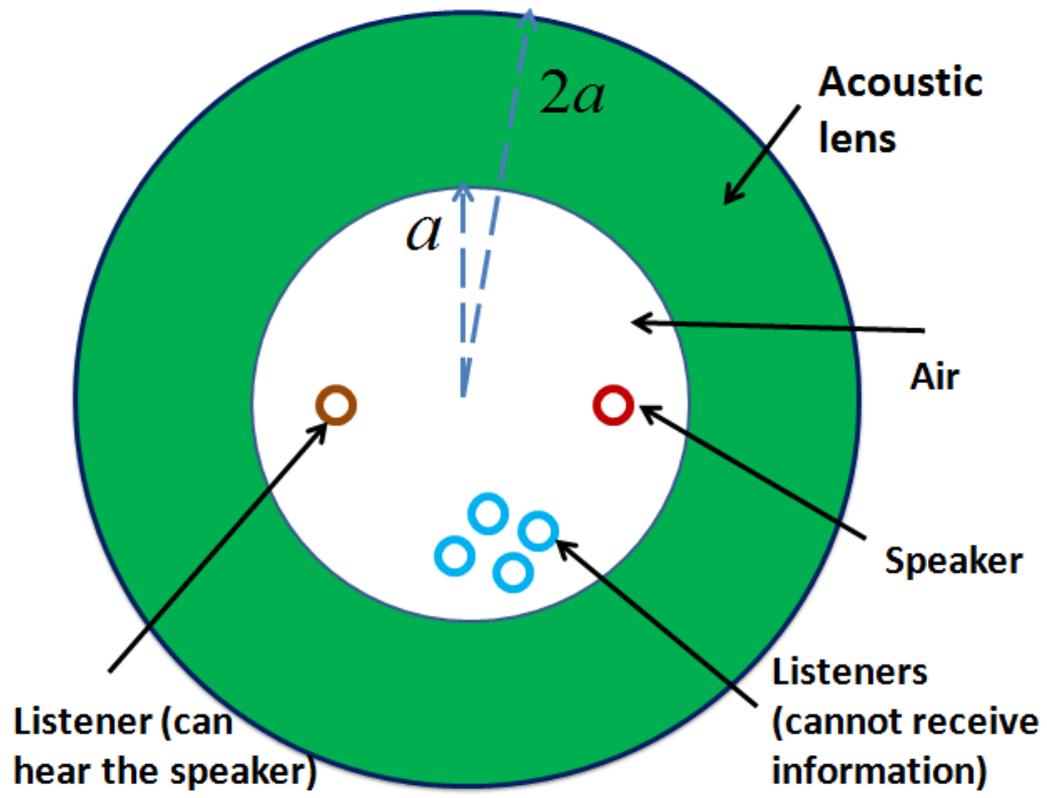

Figure 1| (color online) Basic scheme to achieve a direct point-to-point acoustic communication in air by an AML.

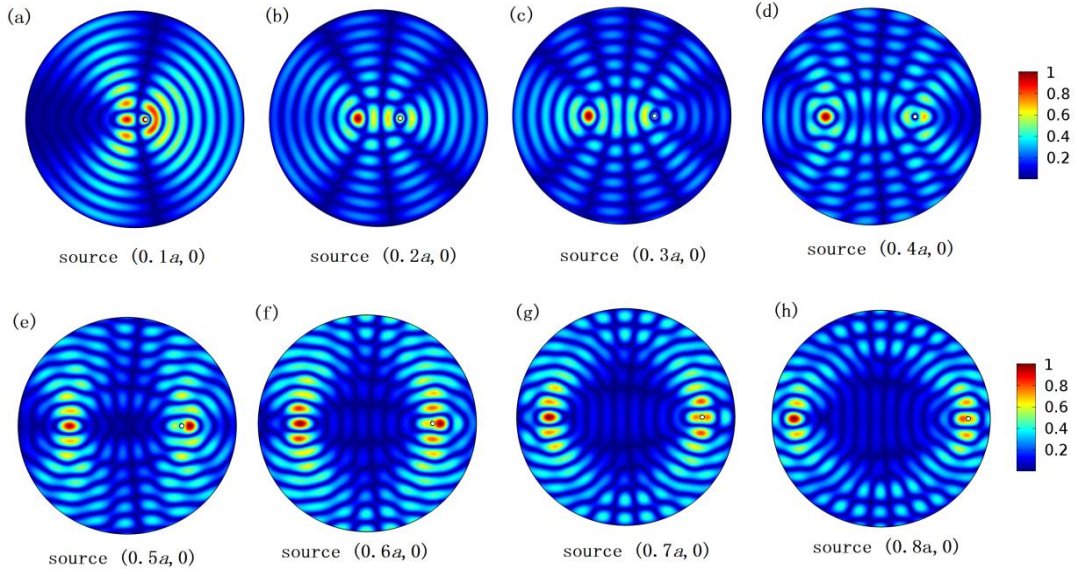

Figure 2| (color online) 2D numerical simulation results. We only plot the normalized acoustic pressure distribution in the center air region of the impedance-matched acoustic inside-out Eaton lens. The mass density and modulus are given by Eqs. (9) and (10), respectively. From (a) to (h), as the position of the acoustic source changes from the center to the edge, the focused acoustic image changes accordingly. We choose $a=1m$, $\rho_0=1.29 kg/m^3$, $v_0=340 m/s$, and wavelength $\lambda_0=0.25m$ in all simulations.

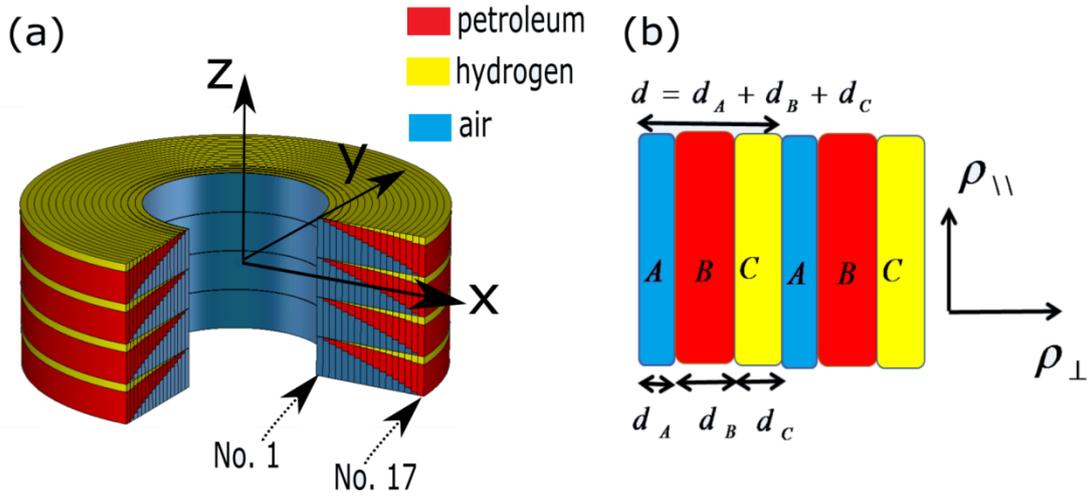

Figure 3| (a) a schematic diagram of our 3D AML composed by air (colored blue), petroleum (colored red), and hydrogen (colored yellow). (b) the effective medium theory of layered media.

| Layer Number | Inner Radius [a] | Outer Radius [a] | Filling Factor of Air f1 | Filling Factor of petroleum f2 | Filling Factor of hydrogen f3 | Effective Mass Density [kg/m3] | Effective Modulus [Pa] |
|---|---|---|---|---|---|---|---|
| 1 | 1 | 1.05 | 0.9885 | 0.0089 | 0.0026 | 1.258143231 | 152895.7874 |
| 2 | 1.05 | 1.1 | 0.9273 | 0.0622 | 0.0105 | 1.196620241 | 160756.766 |
| 3 | 1.1 | 1.15 | 0.868 | 0.1135 | 0.0185 | 1.137673064 | 169086.1866 |
| 4 | 1.15 | 1.2 | 0.8102 | 0.1631 | 0.0267 | 1.08093045 | 177962.2361 |
| 5 | 1.2 | 1.25 | 0.7535 | 0.2114 | 0.0351 | 1.026059433 | 187479.1985 |
| 6 | 1.25 | 1.3 | 0.6975 | 0.2587 | 0.0438 | 0.972755335 | 197752.5007 |
| 7 | 1.3 | 1.35 | 0.642 | 0.3051 | 0.0529 | 0.920732891 | 208925.7394 |
| 8 | 1.35 | 1.4 | 0.5864 | 0.3511 | 0.0625 | 0.869717823 | 221180.7037 |
| 9 | 1.4 | 1.45 | 0.5303 | 0.3969 | 0.0728 | 0.81943819 | 234752.0561 |
| 10 | 1.45 | 1.5 | 0.4733 | 0.4428 | 0.0839 | 0.769614696 | 249949.4891 |
| 11 | 1.5 | 1.55 | 0.4148 | 0.4892 | 0.096 | 0.719948769 | 267192.3453 |
| 12 | 1.55 | 1.6 | 0.3538 | 0.5366 | 0.1096 | 0.670106601 | 287065.9677 |
| 13 | 1.6 | 1.65 | 0.2894 | 0.5856 | 0.125 | 0.619695955 | 310418.0338 |
| 14 | 1.65 | 1.7 | 0.2202 | 0.6369 | 0.1429 | 0.568229773 | 338533.4758 |
| 15 | 1.7 | 1.75 | 0.1436 | 0.6918 | 0.1646 | 0.515064369 | 373477.2031 |
| 16 | 1.75 | 1.8 | 0.0559 | 0.7521 | 0.192 | 0.459284502 | 418835.8181 |
| 17 | 1.8 | 1.85 | 0 | 0.8 | 0.2 | 0.399463682 | 481557.6707 |

Table 1| detailed parameters for each layer in Fig. 3(a).

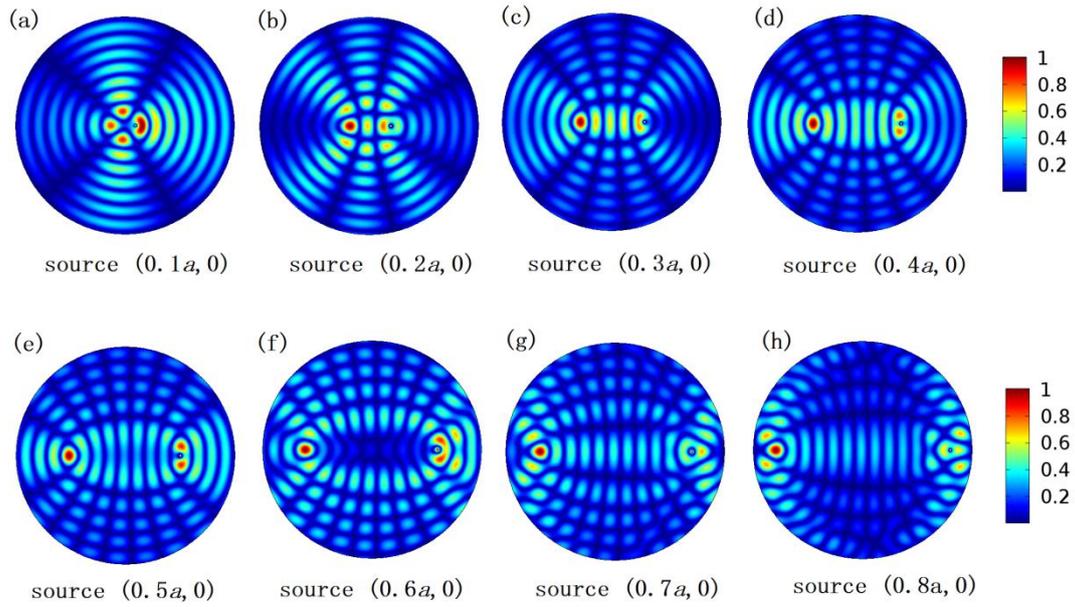

Figure 4| (color online) 2D numerical simulation results. We plot the normalized amplitude of the acoustic pressure distribution in the simplificated AML. The structure and parameters of the simplificated AML are given in Fig. 3(a) and Tab. 1, respectively. From (a) to (f), the position of the acoustic source changes from the center to the edge, its point-to-point image changes accordingly. All other setting are the same as ones used in Fig. 2.